\documentclass[12pt]{article}
\usepackage{epsfig}
\def\neuto{\tilde{\chi}_1^0}
\def\ltap{\ \raisebox{-.4ex}{\rlap{$\sim$}} \raisebox{.4ex}{$<$}\ }
\def\gtap{\ \raisebox{-.4ex}{\rlap{$\sim$}} \raisebox{.4ex}{$>$}\ }

\def\rggt{$R_{\gamma \gamma}$}

\def\mchargo{m_{\tilde{\chi}_1^+}}
\newcommand{\ra}{\rightarrow}
\def\tgb{\tan \beta}
\def\be{\begin{equation}}
\def\ee{\end{equation}}
\def\rgg{R_{\gamma \gamma}}
\newcommand{\np}{Nucl.\,Phys.\,}
\newcommand{\pl}{Phys.\,Lett.\,}
\newcommand{\pr}{Phys.\,Rev.\,}
\newcommand{\prl}{Phys.\,Rev.\,Lett.\,}

\newcommand{\zp}{Z.\,Phys.\,}

\def\Journal#1#2#3#4{{#1} {\bf #2} (#4) #3}

\def\NPB{{\em Nucl. Phys.} B}
\def\PLB{{\em Phys. Lett.}  B} 

\def\EUJ{{\em Eur. Phys. J.} C} 
\def\PRL{\em Phys. Rev. Lett.}
\def\PRD{{\em Phys. Rev.} D}

\def\JHEP{\em Journal of High Energy Physics}
\def\pramana{\em Pramana}

\begin{document}

\begin{titlepage}

\begin{flushright}
                                                   CERN-TH/20002-141\\
                                                   LAPTH-Conf-919\\
                                                   hep-ph/0206311  \\
\end{flushright}

\vspace{0.5cm}
\begin{center}
{\Large

{\bf    Invisible Decays of the Supersymmetric Higgs and Dark Matter}}\\[5ex]
F. Boudjema$^a$\footnote{e-mail:boudjema@lapp.in2p3.fr}
G. B\'elanger$^a$\footnote{e-mail:belanger@lapp.in2p3.fr}
R.M. Godbole$^b$\footnote{On leave of absence from Centre for
Theoretical Studies, Indian Institute of Science, Bangalore,
560 012, India. e-mail:godbole@cern.ch}\\[1.5ex]

{\it a. Laboratoire de Physique Th\'eorique, LAPTH\\}
{\it Chemin de Bellevue, B.P. 110, F-74941 Annecy-le-Vieux,
Cedex, France.}\\
{\it b. CERN, Theory Division, CH-1211, Geneva, Switzerland.}

\end{center}
\vspace{1.0cm}
{\begin{center} 
ABSTRACT

\vspace{2cm}

\parbox{15cm}{ 
We discuss effects of the light sparticles on decays of the
lightest Higgs in a supersymmetric model with nonuniversal gaugino masses at
the high scale, focusing on the `invisible' decays into neutralinos. These 
can impact significanlty the discovery  possibilities of the lightest Higgs 
at the LHC. We show that due to these decays, there exist regions of the 
$M_2-\mu$ space where the B.R.  $(h \rightarrow \gamma \gamma)$ becomes 
dangerously low even after imposing
the LEP constraints on the sparticle masses,  implying a possible 
preclusion of its discovery in the $\gamma \gamma$ channel.  We find that 
there exist regions in the parameter space with acceptable relic density and 
where the ratio ${ B.R. (h \rightarrow \gamma \gamma)_{SUSY} \over
B.R. (h \rightarrow \gamma \gamma)_{SM} }$ falls below 0.6, implying loss 
of signal in the $\gamma \gamma$ channel.
These regions correspond to $\tilde \chi_1^+, \tilde \chi_2^0$ masses
which should be accessible already at the Tevatron. Further we find 
that considerations of relic density put lower limit on the $U(1)$ gaugino mass 
parameter $M_1$ independently of $\mu , \tan \beta$ and $m_0$.
}
\end{center}}

\vfill 
{\begin{center} 
{\it Presented  by R.M. Godbole at}                       \\
{\it Appi2002, Accelerator and Particle Physics Institute} \\
{\it Appi, Iwate, Japan, February 13--16 2002}
\end{center}}

\vfill
\end{titlepage}

\thispagestyle{empty}

\begin{center}

{\Large

{\bf    Invisible Decays of the Supersymmetric Higgs and Dark Matter}}\\[5ex]
F. Boudjema$^a$\footnote{e-mail:boudjema@lapp.in2p3.fr}
G. B\'elanger$^a$\footnote{e-mail:belanger@lapp.in2p3.fr}
R.M. Godbole$^b$\footnote{On leave of absence from Centre for
Theoretical Studies, Indian Institute of Science, Bangalore,
560 012, India. e-mail:godbole@cern.ch}\\[1.5ex]

{\it a. Laboratoire de Physique Th\'eorique, LAPTH}\\
{\it Chemin de Bellevue, B.P. 110, F-74941 Annecy-le-Vieux,
Cedex, France.}\\
{\it b. CERN, Theory Division, CH-1211, Geneva, Switzerland.}

\end{center}
\vspace{1.5cm}
{\begin{center} ABSTRACT \end{center}}
\vspace{-4truemm}
{\small{
We discuss effects of the light sparticles on decays of the
lightest Higgs in a supersymmetric model with nonuniversal gaugino masses at
the high scale, focusing on the `invisible' decays into neutralinos. These 
can impact significanlty the discovery  possibilities of the lightest Higgs 
at the LHC. We show that due to these decays, there exist regions of the 
$M_2-\mu$ space where the B.R.  $(h \rightarrow \gamma \gamma)$ becomes 
dangerously low even after imposing
the LEP constraints on the sparticle masses,  implying a possible 
preclusion of its discovery in the $\gamma \gamma$ channel.  We find that 
there exist regions in the parameter space with acceptable relic density and 
where the ratio ${ B.R. (h \rightarrow \gamma \gamma)_{SUSY} \over
B.R. (h \rightarrow \gamma \gamma)_{SM} }$ falls below 0.6, implying loss 
of signal in the $\gamma \gamma$ channel.
These regions correspond to $\tilde \chi_1^+, \tilde \chi_2^0$ masses
which should be accessible already at the Tevatron. Further we find 
that considerations of relic density put lower limit on the $U(1)$ gaugino mass 
parameter $M_1$ independently of $\mu , \tan \beta$ and $m_0$.
}}


\section{Introduction}
\label{intro} 

\vspace*{0.5truecm}
The importance of the search for  the Higgs particle in the current and 
upcoming collider experiments, the TEV-II, LHC and possibly the next Linear
Colliders, to confirm the crucial features  of the Standard Model (SM) of
the fundamental particles and interactions among them, can not be 
overemphasised\cite{Boudjema:2001ni}. Further, supersymmetry (SUSY)
is one of the most attractive ways to go beyond the SM and provide a cure for 
one of its most serious theoretical ills {\it viz.} the hierarchy problem in the
scalar sector\cite{Kaul:wp}. Therefore, looking for the evidence of the 
extended Higgs sector of the supersymmetric model also forms a very important 
part of the planned research program of the current and future accelerator
experiments. In this talk we discuss some aspects of the effect that the 
supersymmetric partners (the sparticles) can  have on the decays of the 
lightest neutral scalar present in the Higgs sector of the supersymmetric
theories, with special emphasis on those SUSY models where the gaugino
masses are not unified at the high scale.  The plan of this talk is as 
follows. We first summarise a few relevant facts about the expected Higgs 
spectrum in the supersymmetric models as well as a few details about the 
SM Higgs and the lightest neutral SUSY scalar, such as the theoretical as well 
as the current experimental bounds on its mass etc. We then discuss the 
effect of light superpartners on the couplings and the decay of the Higgs, 
notably the `invisible' decay into a pair of neutralinos and its implications 
for the Higgs search at the LHC. We then examine the range of values predicted 
for B.R. $(h \rightarrow {\rm invisibles})$ once the current experimental 
constraints on the Dark Matter (DM) are implemented. We then end with a few 
remarks about  probing at the Tevatron the region of the $M_2 - \mu$  parameter
space, where the B.R. $ (h \rightarrow {\rm invisibles})$ is substantial, yet
the DM constraint is satisfied, as well as about looking for such an `invisible'
Higgs at the LHC through its associate production with a $W/Z$.

\section{SM and MSSM Higgs: Masses and Couplings}
\label{summary}
The SM has precise predictions  for the couplings of the $h$ but can predict 
only limits on its mass.  According to these limits given by
the consideration of vacuum stability and triviality, using 2-loop 
RGE equations~\cite{hambye} the mass of the SM Higgs should 
lie in the range $160 \pm 20$ GeV if there is no new physics between the EW
scale and the Planck scale. The high precision measurements of the $Z$-boson
properties and those of $M_W$ at LEP, of $\sin^2\theta_w$ at SLD, as
well as the measurements of $m_t,M_W$ at the Tevatron,
all put together constrain the Higgs mass substantially and give an upper 
limit on its mass of $196$ GeV at $95 \%$ c.l.\cite{LEPC}. Thus these indirect
measurements of the Higgs mass prefer a light Higgs and the  consistency of 
this indirect upper  limit with the above mentioned range of $160 \pm 20$ GeV, 
is very tantalising.  Very general theoretical arguments about  the 
'naturalness' requirements also indicate that the Higgs mass be small and 
of the order of the
EW scale \cite{barbieri}.  Furthermore, lack of any `direct' experimental 
evidence for the Higgs in the process $e^+e^- \rightarrow Z h$  puts a 
limit \cite{hlimit} $m_h \gtap 113$ GeV, with a hint of a signal for a Higgs 
with mass close to this upper limit. Thus in the SM clearly a light Higgs is 
preferred, both experimentally and theoretically.

In the Supersymmetric theories the situation is not any different. 
These theories have to  have two Higgs doublets for reasons of anomaly
cancellations as well as to give mass to both the up and down-type
fermions.  Of the three neutral scalars $h,H$ and $A$ the first two 
are $CP$ even and the last one is $CP$ odd in the Minimal Supersymmetric 
Standard Model (MSSM).  Supersymmetry keeps the mass of the lightest SUSY 
scalar $m_h$ low `naturally' for symmetry reasons. In these theories it is 
actually predicted in terms of $M_Z,$ and the gauge coupling, being  bounded 
from above by $M_Z$ at the tree level. Large loop corrections due to the 
heavy top modify this upper limit to $\sim 135$ GeV \cite{heinemeyer} in the 
MSSM and to $\sim 165$ GeV in the NMSSM\cite{kolda,spanish,pandita}. These 
upper limits are really quite robust and have very little dependence on most 
of the minimal SUSY model parameters, except on the  trilinear parameter $A_t$, 
$\mu, \tan \beta$ through the $L-R$ mixing in the stop sector and the squark 
mass term for the top squarks. The direct experimental lower limits in the case 
of the MSSM, are $91.0$ GeV \cite{lepsush} for the the $CP$ even Higgs  and 
$91.9$ for the $CP$ odd Higgs.

Therefore the search for a light SM Higgs and the lightest SUSY Higgs 
({\it i.e.} $m_h < 2 M_W$) deserves a special emphasis while assessing the 
capabilities of any collider, present or future.  Although, it is true 
that a light Higgs, if found,  can not be taken as a 'proof' of Supersymmetry, 
it is certain to boost our belief in weak scale SUSY. It is also clear 
that a discussion of the effect of sparticles on Higgs searches  is 
also quite crucial. At the  LHC the dominant mode of production of the
Higgs is through its coupling to the gluons induced by the diagrams shown in 
\begin{figure}[ht] 
\centerline{ 
\includegraphics[scale=0.55]{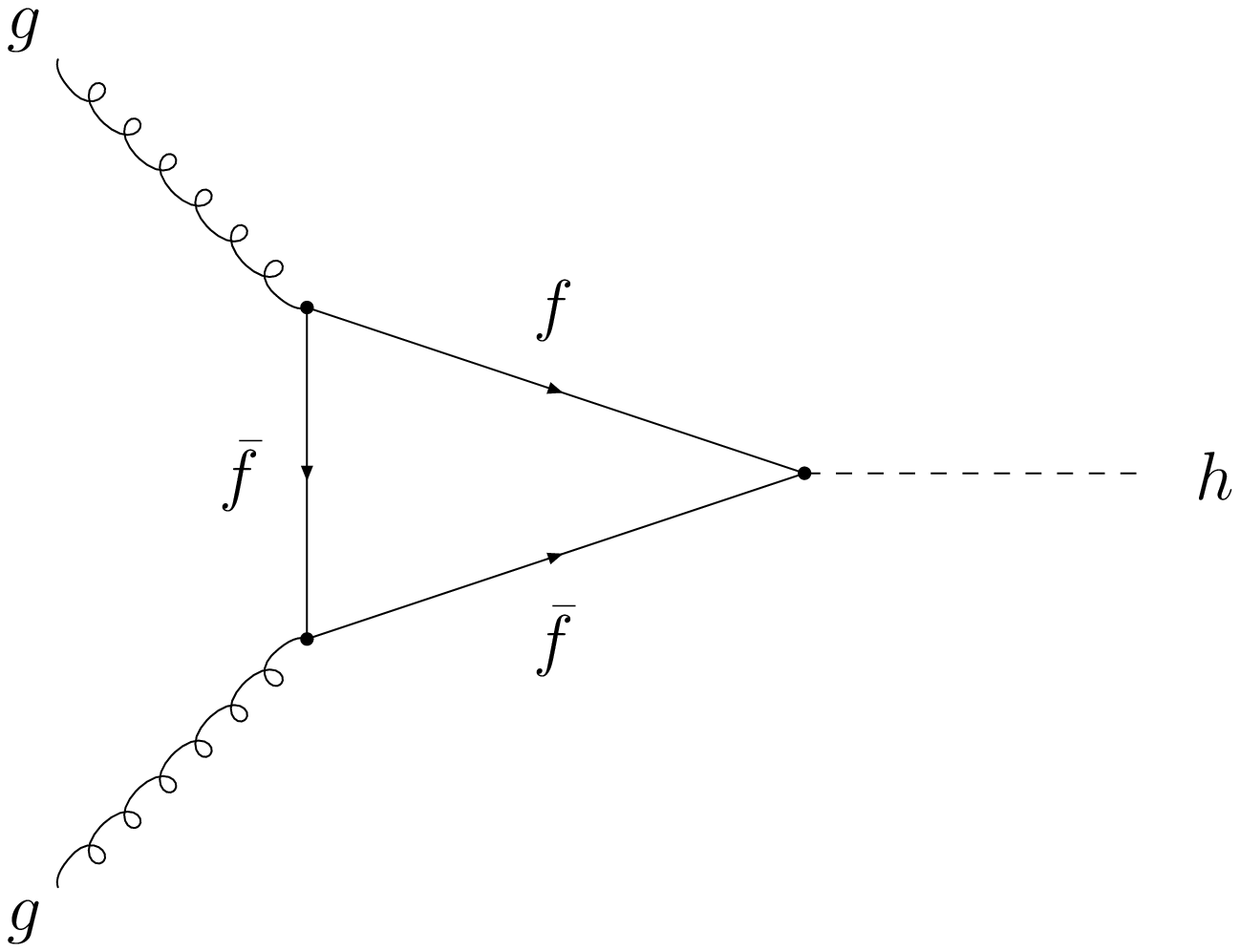} 
\hspace{-1cm}
\includegraphics[scale=0.55]{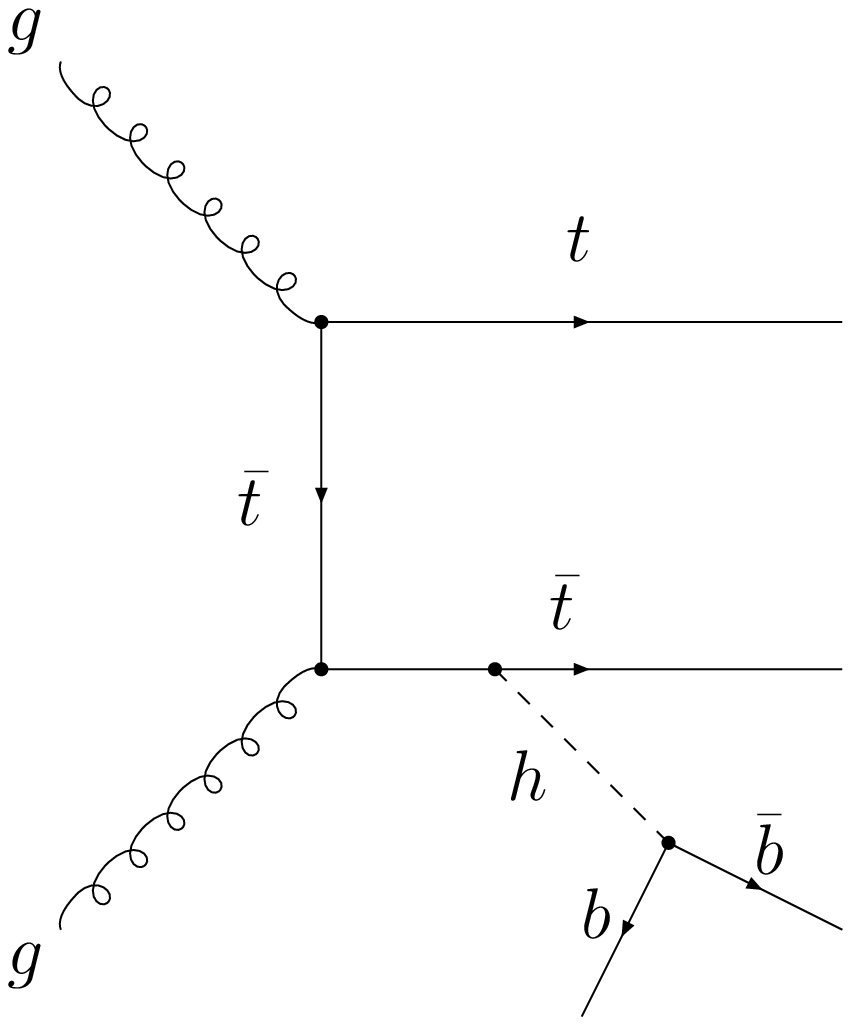}
} 
\caption{\label{ggh} 
{\em Production processes for the $SM$ Higgs in $gg$ collisions}}
\end{figure}
the left panel of Fig.\ref{ggh}. This coupling is dominated by the
contribution of the $t$ quarks in the loop.  For the mass range 
$m_h < 2 M_W$,  the one we are interested in this discussion, the  decay mode 
that can be used mostly for the search of the $h$ in this inclusive 
production mode is $h \rightarrow \gamma \gamma$. This coupling is also 
loop induced and the corresponding diagrams in the SM are shown in
\begin{figure}[ht] 
\centerline{ 
\includegraphics[scale=0.55]{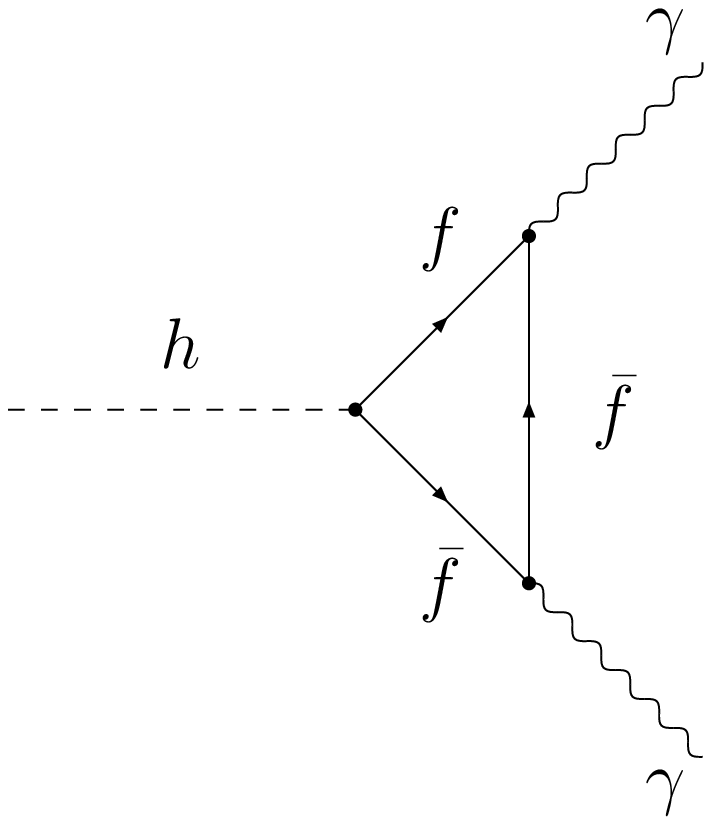} 
\includegraphics[scale=0.55]{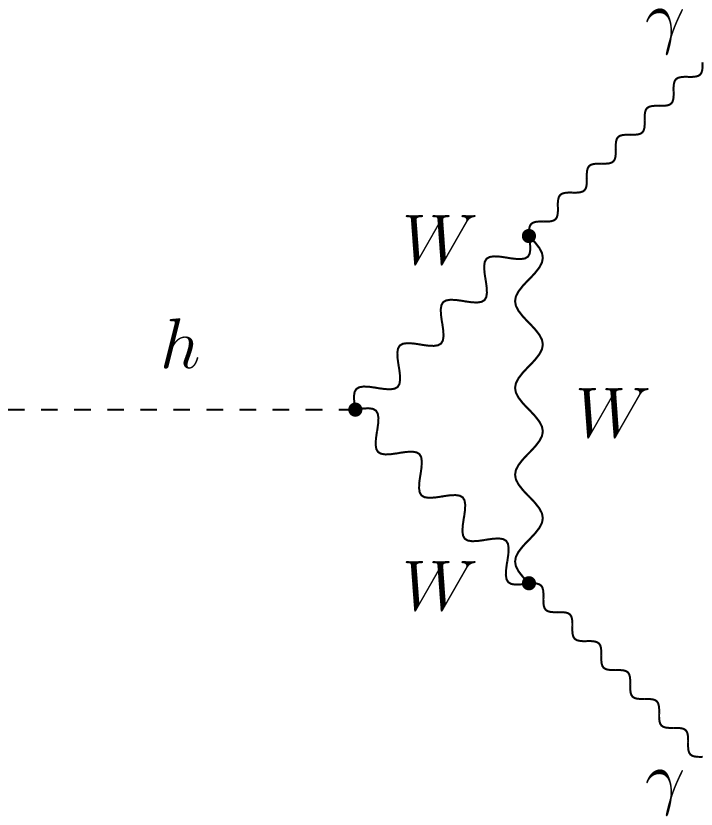}
} 
\caption{\label{hggsm}{\em Loop diagrams giving rise to the 
$h \gamma \gamma$ coupling in the SM.}}
\end{figure}
Fig.\ref{hggsm}.  This decay receives the dominant contribution from 
$W$ loops. Thus for the inclusive production process shown in the left 
panel of Fig.\ref{ggh} we have,  
\be
\sigma (p p \rightarrow h) B.R. (h \rightarrow \gamma \gamma) 
\propto B.R. (h \rightarrow \gamma \gamma ) \times B.R. (h \rightarrow g g).
\label{eq:1}
\ee
Another good signal for the $h$ at the LHC is via  the associated 
$t \bar t h$ production depicted in the right panel of Fig.\ref{ggh}. In this 
case due to the $t \bar t$ quarks present along with the $h$ in the final state,
one can use the dominant $h \rightarrow b \bar b$ decay mode for the search,
the final state consisting of $t \bar t b \bar b$. In this case the search 
channel does not depend on the branching ratio of the $h$ into 
the  $\gamma \gamma$ or the $gg$  channel but does depend on 
B.R. ($h \rightarrow b \bar b$).
Thus the decays which play an important role in the determination of the
search possibilities and reach for a light neutral scalar at the hadronic
colliders are the tree level decay $h \rightarrow b \bar b$ and the loop
induced one $h \rightarrow \gamma \gamma$.

In case of the SUSY Higgs, its couplings  depend on some of the parameters of 
the SUSY model {\it viz.} $m_A, \tan \beta$ and $\mu$. For $m_A \gg M_Z$ the 
tree level couplings of the $h$ to the SM fermions and the gauge bosons are 
very close to that in the SM, in this so called `decoupling limit'. The loop
induced $gg$ and $\gamma \gamma$ couplings which affect the production through 
the $gg$ mode and detection  through the $\gamma \gamma$ mode respectively at 
the hadronic colliders, receive additional contributions from the loops 
containing the charged sparticles which have substantial coupling to the $h$,
{\it viz.}  the ${\tilde t}_1, {\tilde t}_2$, the charginos
${\tilde \chi_1}^\pm, {\tilde \chi}_2^\pm$ and the charged Higgs $H^\pm$. 
These are shown 
\begin{figure}[ht] 
\centerline{ 
\includegraphics[scale=0.60]{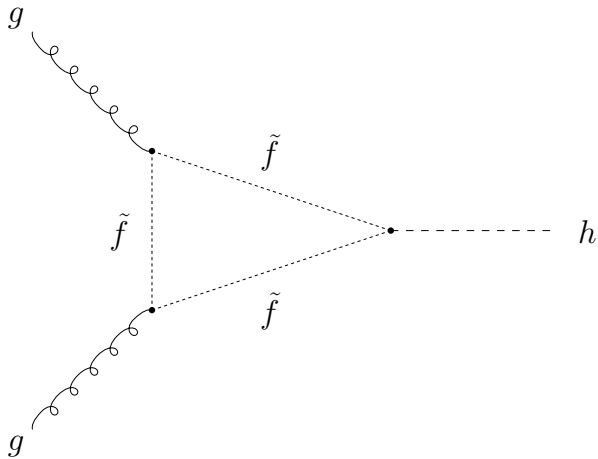} 
} 
\caption{\label{ggsh} 
{\em Additional contributions to the $ggh$ coupling for the SUSY Higgs h}}
\end{figure}
\begin{figure}[ht] 
\centerline{ 
\includegraphics[scale=0.60]{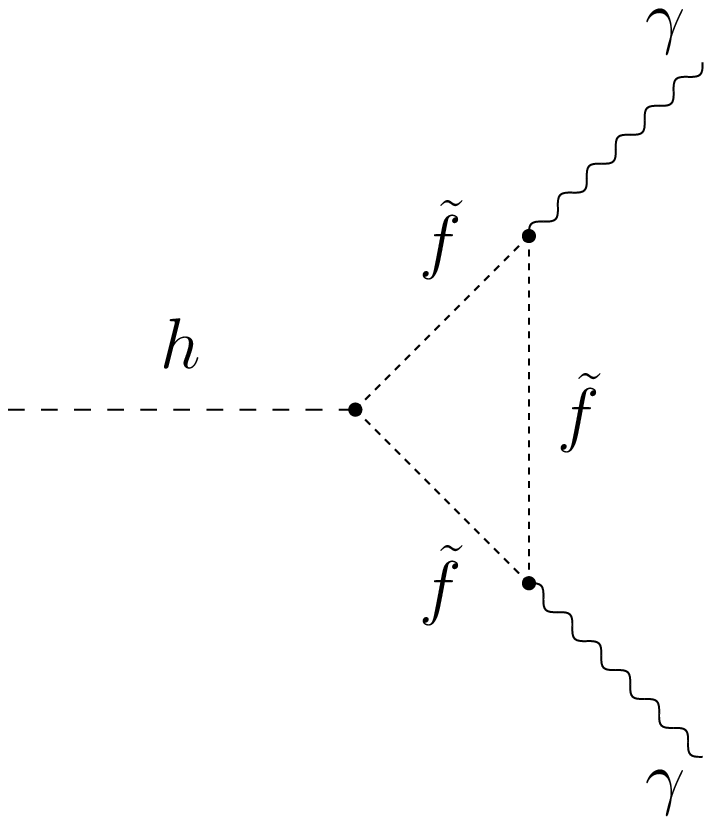} 
\includegraphics[scale=0.60]{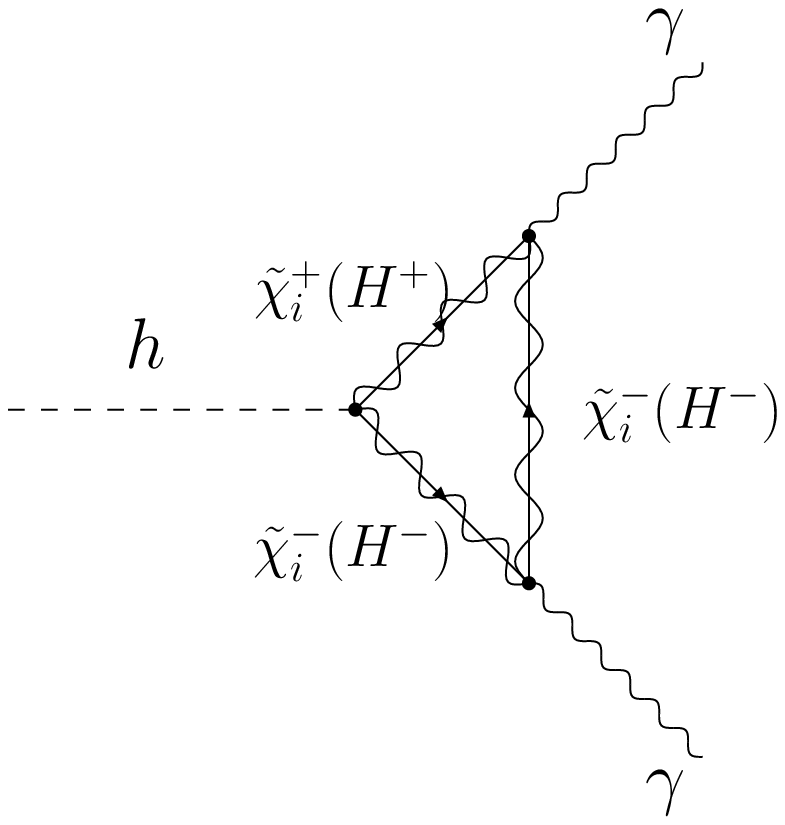}
} 
\caption{\label{hsgg}
{\em Additional sparticle loop contributions to $h\gamma \gamma$ coupling
for the SUSY Higgs}}
\end{figure}
in Figs.~\ref{ggsh},\ref{hsgg} respectively.  For light sparticles,
of course being consistent with the non observation at LEP, these effects
can be large. Particularly strong is the effect of the light top squarks, 
$\tilde t_1, \tilde t_2$, on the $ggh$  coupling. For comparable $t$ and 
$\tilde t_i$ masses and large mixing 
between the left and right chiral top squarks, the $f$ and $\tilde f$ 
contributions interfere destructively and can cause a decrease in the 
B.R. $(h \rightarrow gg)$ lowering the production cross-section thereby.
The decay $h \rightarrow \gamma \gamma$ also receives contribution from loops
containing the charged particles and sparticles with only the EW couplings 
{\it viz.}, the ${\tilde \chi}_1^\pm$ and the $W$'s, along with that from the 
top quarks/squarks. This actually increases  the B.R. w.r.t. expectation  in 
the SM, rising with increasing mixing in the $L-R$ sector for the $\tilde t$.
Further the B.R. $(h \rightarrow \gamma \gamma)$ and B.R. $(h \rightarrow g g)$
are also affected by the decays of the  Higgs $h$ into final states containing
sparticles. In view of the current LEP bounds the only possibility still 
allowed is the decay of $h$ into a pair of neutralinos 
${\tilde \chi}_l^0 {\tilde \chi}_m^0$ \cite{djouadi}.  
The decay $h \ra {\tilde\chi}_1^0 {\tilde \chi}_1^0$ renders the 
Higgs invisible and in addition reduces the branching ratio of the $h$ in 
both the $\gamma \gamma$ and the $b \bar b$ channel, relative to the values 
expected in the SM thereby reducing the significance of these useful channels. 

\section{Effect of light stops on $h$ production/decay and the LHC observables}
The figure of merit at the LHC for the search of a light Higgs $h$ is 
the L.H.S. of Eq. \ref{eq:1} or the corresponding quantity for the 
$ b \bar b$ final state with the $t \bar t h$ associate production.
Hence the effect of sparticles on the light Higgs search at the LHC can be 
best assessed by  studying the ratio
\be 
R_{gg\gamma \gamma} = {\sigma (gg  \rightarrow h) 
B.R.(h \rightarrow \gamma \gamma)_{SUSY} \over
\sigma (gg  \rightarrow h) B.R. (h \rightarrow \gamma \gamma)_{SM}},
\label{rglglgg}
\ee
as well as similar ratios of branching ratios for the SUSY Higgs
and the SM Higgs,  $R_{\gamma \gamma}$ for B.R. $(h \rightarrow \gamma \gamma)$ 
and $R_{b \bar b}$  for B.R. $(h \rightarrow b \bar b)$.  
Effect of the light top squarks, on these ratios and hence on search of the 
SUSY Higgs at the LHC, with all the other sparticles being 
heavy~\cite{djouadi1,kileng,belbous} as well as that of the light 
\begin{figure}[htb] 
\centerline{ 
\includegraphics[scale=0.70]{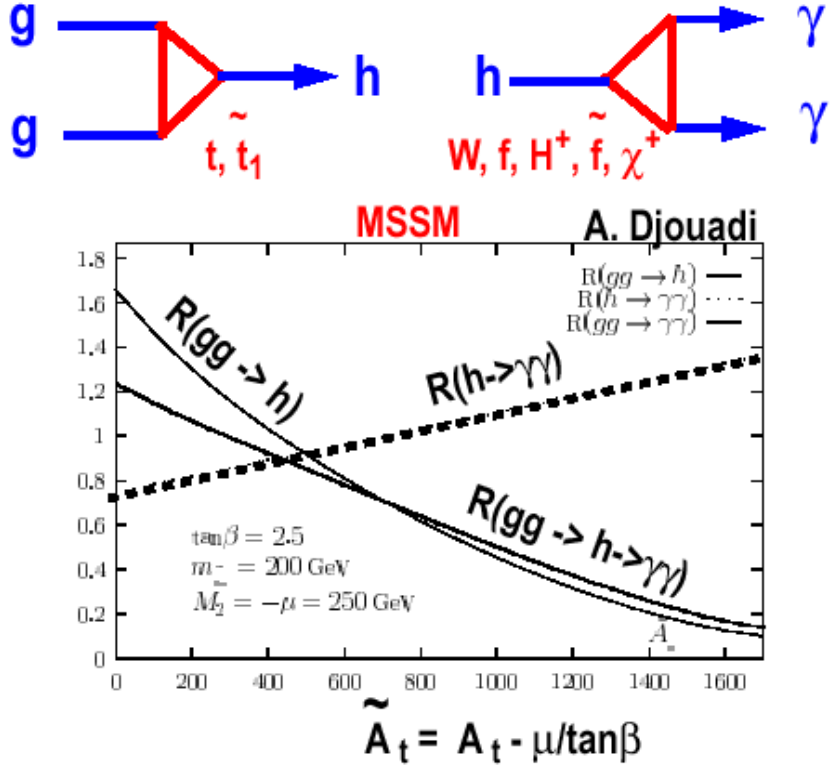} 
}
\caption{\label{abdelst}{\em The ratio  $R_{gg\gamma \gamma},
R_{\gamma \gamma}$ and $R_{gg \rightarrow h}$ as a function of
$A_t, \mu$ and $\tan \beta$, for light top squarks \protect\cite{djouadi1}}}
\end{figure} 
${\tilde \chi}_l^0, {\tilde \chi}_l^\pm$ \cite{belbour} has been studied in 
detail. The analyses show that for $m_{\tilde t_1} \simeq m_t$ and 
large $L-R$ mixing, sensitivity to the light $h$ at LHC can be completely lost. 
This is depicted in the 
\begin{figure}[htb]
\centerline{
\includegraphics[height=9cm]{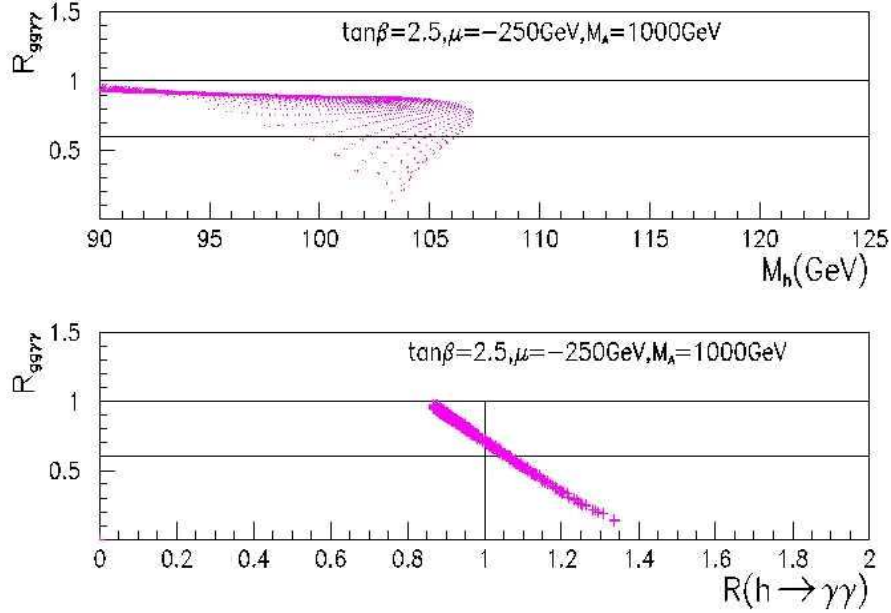}}
\caption{\label{sf03xv} {\em Ratio $R_{gg\gamma\gamma}$ as a function of 
$R_{\gamma \gamma}$ and $m_h$ \protect\cite{belbous}. The values of various
parameters are indicated in the figure.}}
\end{figure}
Figs.~\ref{abdelst} and \ref{sf03xv} taken from Refs.\cite{djouadi1} and 
\cite{belbous} respectively. As one can see from these figures, for large stop 
mixing the ratio $R_{gg\gamma \gamma}$ falls 
below $0.6$  thus losing the signal for the $h$ in the inclusive
$\gamma \gamma$  channel. The choice of $0.6$ is 
arrived at by taking the possible level of significance somewhere between the
ATLAS \cite{atlas_tdr} and CMS simulations \cite{cms}. It was 
shown \cite{belbous} that should the loss of signal be due to the light 
top squark, luckily a viable signal will still exist in the ${\tilde t}_1 
{\tilde t}^*_1 h$ channel, along with the $t \bar t h$  channel mentioned 
above. It should be added here that analysis of optimisation of the search 
for the top squark in this mass range at the LHC is still not done.

\section{Effect of light chargino/neutralinos on the SUSY Higgs production 
and decay.}
In view of the LEP bounds \cite{lepsusy} ($ m_{{\tilde \chi}_1^+} > 103 $
GeV), the only effect that the ${\tilde \chi}_1^+$ can have on the Higgs 
widths and the couplings is through the loop effects on  $h\gamma \gamma$  
coupling.  On the other hand the light neutralinos open up a new channel for 
the $h$ decay and  thus can affect the branching ratios into the 
$\gamma \gamma$ and $b \bar b$ channel.  Since for the mass range of the $h$ 
we are interested in, $h \ra b \bar b$  is the only dominant decay mode, 
we have 
\be
R_{\gamma \gamma} \simeq R_{b \bar b} \simeq 1 - B.R. (H \ra {\tilde \chi}^0_1 
{\tilde \chi}^0_1)
\ee
An increasing branching ratio for the channel $h \ra {\tilde \chi}^0_1 
{\tilde \chi}^0_1$
thus causes a depletion into the $\gamma \gamma $ and $b  \bar b$ channel,
with respect to the SM values. Since we consider the case of heavy 
squarks, the  production rate of the $h$ in the inclusive channel 
$p \bar p \ra ggX \ra h X$ is not affected. Since the chargino/neutralino sector
is completely defined in terms of the SUSY breaking $SU(2)$ and $U(1)$ gaugino 
masses $M_1, M_2$ in addition to $\tan \beta$ and $\mu$, one can study these 
effects as a function of these parameters. The Higgs mass $m_h$ depends 
on $A_t, m_A$ in addition.  Of course, under the assumption of unified 
gaugino masses at 
high scale, $M_1 \simeq 0.5 M_2$ at the EW scale and thus the number of 
independent parameters is reduced by one.  For our studies \cite{belbour} 
we chose moderate $\tan \beta$ and large $A_t$, to maximise $m_h$ and $\tilde \chi_1^0 \tilde \chi_1^0 h$ coupling and still have light enough 
$\tilde \chi_1^0$ , thus 
enhancing the possibility of direct decays of the $h$ into a neutralino pair. 
Further, if one also assumes unification of the gaugino masses at  high scale,
then the observed experimental limits on the $m_{{\tilde \chi}_1^\pm}$ of $\sim 
103 $ GeV implies a limit on the $m_{{\tilde \chi}_1^0}$ of about $60$ GeV
reducing the phase space for the `invisible' $h$ decay into a neutralino 
pair. The current LEP bounds on the masses of all other sfermions make
the decays of $h$ into a $\tilde f \tilde f^*$  pair impossible.
\begin{figure}
\begin{center}
\includegraphics[width=14cm,height=15cm]{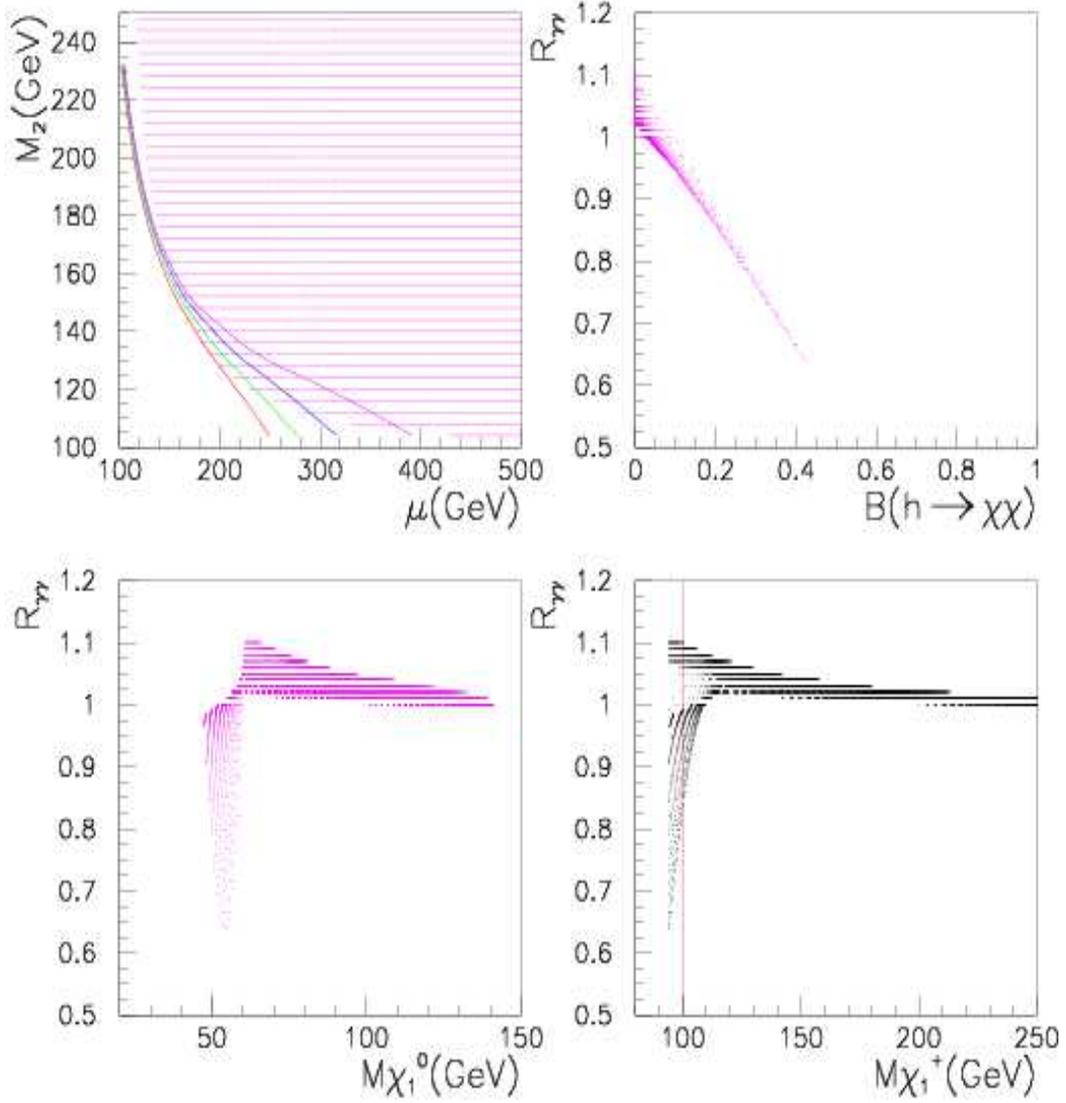}
\caption{\label{tan5at2400} {\em a) Contour plot of $Br(h\ra
\neuto \neuto)=0.1,0.2,0.3,0.4$ (from right to left respectively)
in the $M_2-\mu$ plane. The shaded area is the allowed region.b)
Correlation between \rggt and $Br(h\ra \neuto \neuto)$  c)
Variation of $R_{\gamma\gamma}$ with  the mass of the LSP
$M_{\tilde \chi_1^0}$ and d) mass of the chargino $m_{\tilde \chi_1^+}$. The
vertical line corresponds to $\mchargo=100$GeV. All plots are for
$\tgb=5$, $M_2=50-300$GeV, $\mu=100-500$GeV and $A_t=2.4$TeV \/.}}
\end{center}
\end{figure}
Fig.\ref{tan5at2400} shows first our results where we assume the
gaugino mass unification at high scale. Panel $(a)$ shows the region in 
the $M_2 - \mu$ plane which is allowed by the experimental  limit on the 
$\tilde \chi_1^+$ mass along with contours of B.R. 
$(h \ra \tilde \chi^0_1 \tilde \chi^0_1)$. We see that largish values for 
this branching ratio  are allowed only  close to 
the edge of the allowed region in the $M_2 - \mu$ plane, consistent with the
general argument presented above. The  remaining panels show correlation of
\rggt\ with B.R. $(h \ra {\tilde \chi}_1^0 {\tilde \chi}_1^0)$, the mass of the
chargino and the LSP. We see clearly that for the case of the light 
${\tilde \chi}_1^\pm, {\tilde \chi}_1^0$, the dominant effect on \rggt\ 
is through the `invisible' decays of the $h$ once the LEP constraints on the 
$m_{{\tilde \chi}_1^+}$ are imposed. The reduction is close to the dangerous 
value of $0.6$ - $0.7$ over a very small region of the $M_2-\mu$ space 
in this case.  This can be easily understood by looking at the conditions
that maximise the  B.R. $(h \ra {\tilde \chi}_1^0 {\tilde \chi}_1^0)$. Recall,
$\neuto$ is a  mixture of gaugino/higgsino given by $\neuto = N_{11} \tilde B
+ N_{12} \tilde W_3 + N_{14} \tilde H_1^0 + N_{13} \tilde H_2^0$ and the
$h \neuto \neuto$ coupling $C_{h \neuto \neuto} \propto (N_{12} - \tan \theta_w N_{11}) \times (N_{14} \sin \beta - N_{13} \cos \beta) $. To maximise this
coupling  the $\neuto$ needs to have both the Higgsino and the Gaugino 
components at a sizable level. For a light LSP  and hence a small $M_1$ this
requires small $\mu$. For  $h \ra \neuto \neuto$ to be possible,
we need further $m_{\neuto} \ltap 65$ GeV. Since we also have to impose,
$m_{{\tilde \chi}_1^+} > 103 $ GeV, the values of $\mu$ are bounded from 
below.  Hence, the region where these conditions are satisfied is rather small.
Thus, apart from the degenerate case where $m_{{\tilde \chi}_1^+} \simeq 
m_{\tilde \nu} $ and hence the LEP constraints on the chargino mass are
not applicable, for the case with unified gaugino masses at the high scale 
the `invisible' Higgs decays can not cause a big reduction in \rggt\ and hence
does not pose a big danger to the $h$ search.

However, even for mSUGRA  the unification of gaugino masses at  high 
scale is true only  for the case where the kinetic term for the gauge 
superfields is minimal \cite{md1}. Non-universal gaugino masses are expected 
also in models with Anomaly mediated SUSY breaking (AMSB) \cite{AMSB} or moduli 
dominated SUSY breaking \cite{mod}. In general therefore, we can expect
$M_1 = r M_2$  with $r \ne 0.5$ at the EW scale. We therefore study the 
effect of relaxing the assumption of universal gaugino masses on the 
`invisible' decays of the $h$.  A ratio $r$ between the two gaugino masses
at the EW scale needs, 
\be
M_1 = 2 r M_2,
\label{m1m2}
\ee 
at the GUT scale. Since we want to explore regions of parameter space which 
maximise the B.R. $(h \ra \neuto \neuto)$, we necessarily need  
$\neuto$  lighter than the ones allowed in the case of universal gaugino 
masses and hence $r < 1$.  One should note here  that most of the models 
mentioned above give $r > 1$.  So our choice of $r <1$ is to be treated as 
completely phenomenological. 
\begin{figure}[htbp]
\begin{center}
\includegraphics[width=16cm,height=18cm]{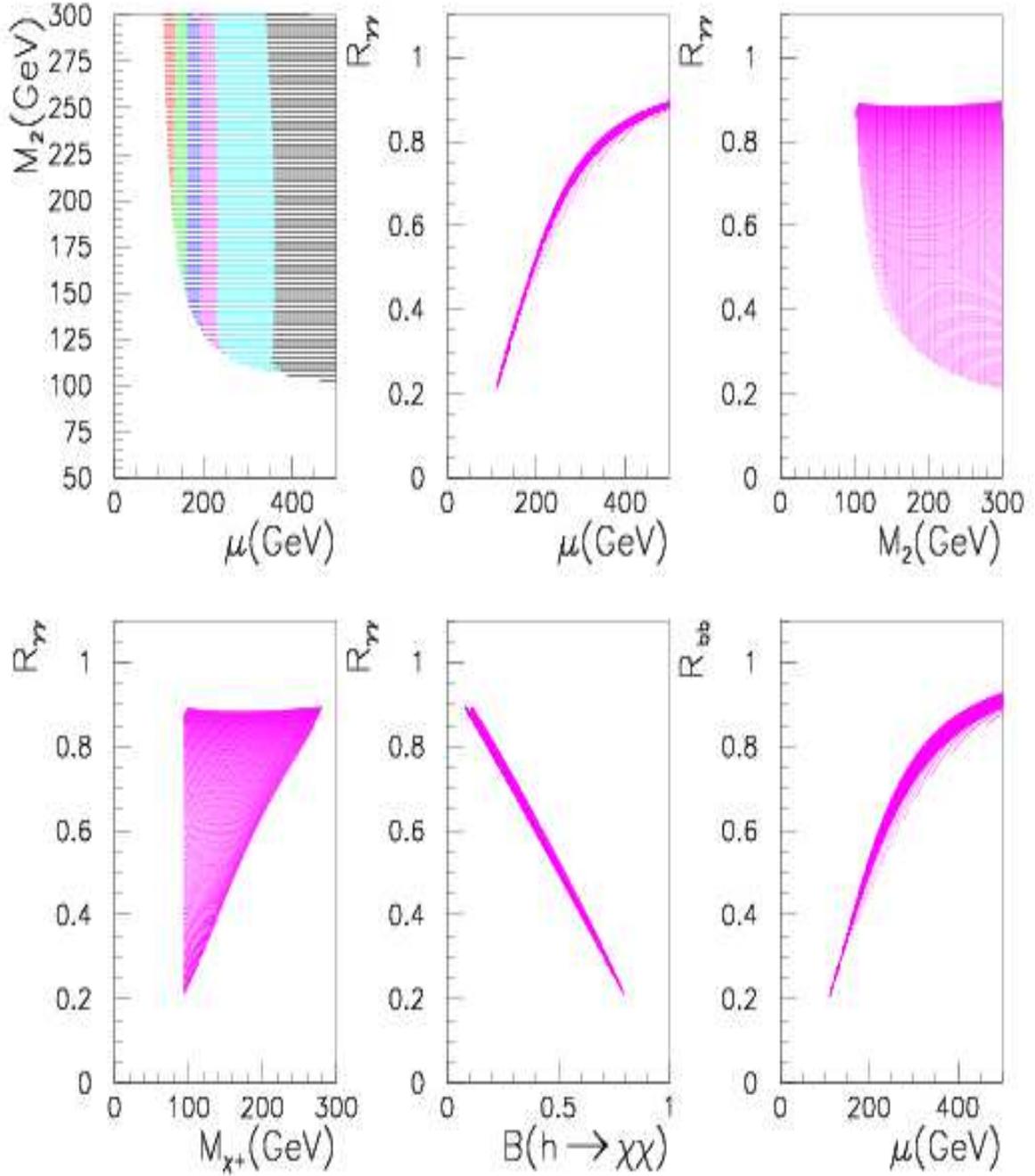}
\caption{\em Effects of neutralinos from
$M_1=M_2/10$ with $\tgb=5$ and $A_t=0$ with heavy selectrons. In
all the plots, scans are over $M_2=50-300$GeV, $\mu=100-500$GeV.
From left to right and top to bottom a) Density plot for \rggt  in
the allowed $M_2-\mu$ plane. The different shadings correspond to
$ .3 <\rgg <.4$ (left band) to $ .8 <\rgg <.9$ (right band). b)
Variation of \rggt with $\mu$ c) with $M_2$ d) with the mass of
the chargino $m_{\chi_1^+}$. e) Correlation between \rggt and the
branching into LSP. f) Variation of $R_{b \bar b}$ with $\mu$\/.}
\label{no-unif-heavysl}
\end{center}
\end{figure}
Fig.~\ref{no-unif-heavysl} shows B.R. $(h \ra \neuto \neuto)$ as a function of 
$M_2$ and $\mu$ as well as correlations between \rggt with 
$\mu, m_{{\tilde\chi}_1^+}, M_2$ and B.R. 
$(h \ra {\tilde \chi}^0_1 {\tilde \chi}^0_1)$ for $r=0.1$. Note that 
one needs to reinterpret the LEP allowed regions in the $M_2 -\mu$
plane for our chosen value of $r = 0.1$.  For these plots we have taken the
selectrons to be heavy just like the squarks. We see, indeed that there exist 
now large regions at low $\mu$  where \rggt\ dips below the dangerous limit of 
$0.6$.  The plots also clearly show that the dip in \rggt\ comes essentially 
from the opening up of the decay channel $h \ra \neuto \neuto$. The last 
panel $(f)$ in the figure also shows that the same of course causes a 
depletion in $R_{b \bar b}$ and further $R_{b \bar b} \simeq  
R_{\gamma \gamma} = 1 - B.R. (h \ra \neuto \neuto)$. Thus the significance 
of the reach at the LHC in the inclusive channel as well as the 
$t \bar t h$ channel is affected by the `invisible' decays of the $h$ quite 
substantially over a large region in the $M_2 - \mu$ space once we 
allow $r \ne 1$.  The effects are more modest for larger values of $\tan \beta$
as the rise  in the $\neuto$ mass with $\tan \beta$ is much more than the
inrease  in the value of $m_h$.

\section{Light ${\tilde \chi}_1^0$ and the cosmological relic density}
Thus we see that models with nonuniversal gaugino masses can allow 
`invisible' decays of the $h$ into a pair of LSP's, at a level which can 
bring down the branching ratio of the $h$ into the discovery channels of 
$\gamma \gamma$ and $b \bar b$  to low enough values threatening to preclude 
its discovery at the LHC. We also saw that this basically needs a light 
$\neuto$.
However, such a stable, light $\neuto$ has  also cosmological implications.
Such a WIMP $\neuto$, is an ideal Dark Matter (DM) candidate. The relic
density of the DM is decided by the annihilation cross-sections 
$\sigma (\neuto \neuto  \ra \ f^+ f^-)$.  Some of the diagrams
contributing to these are shown in 
\begin{figure}[ht] 
\centerline{ 
\includegraphics[scale=0.50]{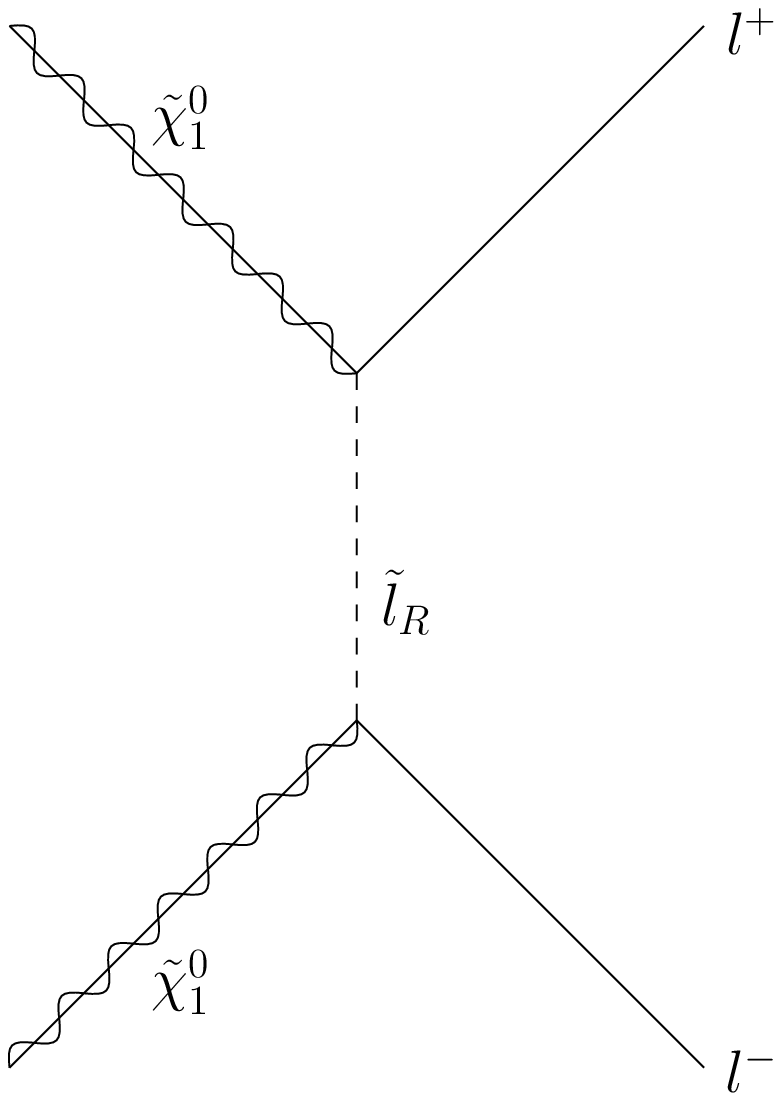} 
\includegraphics[scale=0.55]{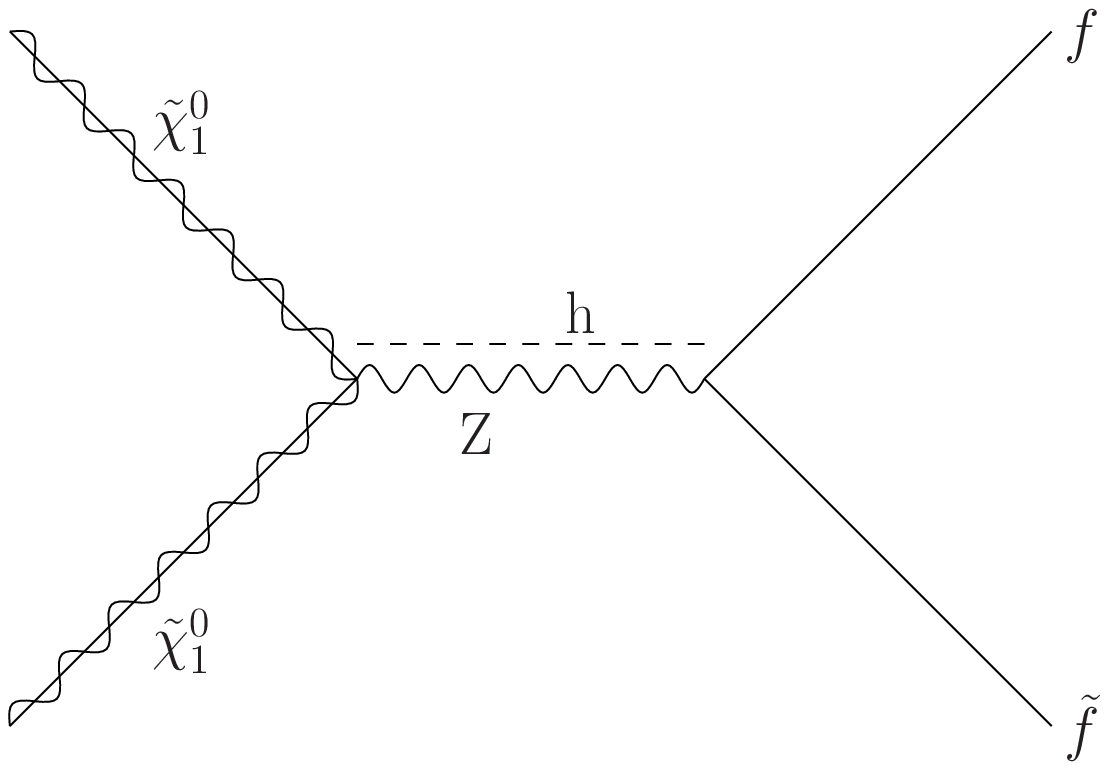}
} 
\caption{\label{dmlzh}
{\em Annihilation of a $\neuto \neuto$ pair via a $\tilde l_R, Z$ or $h$.}}
\end{figure}
Fig.~\ref{dmlzh}. A light $\neuto$ such as the one we are looking at
which is  mostly a Bino, gets largest contributions to the annihilation 
cross-sections via  diagrams involving a light $\tilde l_R$. For somewhat 
heavier  $\neuto$ which can annihilate through a $h/Z$  the relic 
density is reduced very effectively when the exchanged $h/Z$ is on mass shell.
A code \cite{fbp} which includes all the coannihilation channels as well as 
tackles all the $s-$ channel poles and threshold effects is used to calculate
the relic density for the sparticle mass spectrum obtained with $r=0.1,0.2$ 
and with a common scalar mass (defined at the GUT scale) for all the 
three generations of the light and left chiral sleptons and taking all 
the squarks to be heavy. The squarks can be much heavier with the
same common scalar mass at the GUT scale due to much larger $SU(3)$ 
contributions that the squark masses receive. Various observations \cite{dm}
suggest that $0.1 < \Omega h^2 < 0.3$, where $\Omega$ is the fraction of 
the
critical energy density provided by the neutralinos and $h$ is the Hubble
constant in units of $100$ km s$^{-1}$Mpc$^{-1}$. Our choice of the  upper 
limit  is indeed very conservative in view of the recent measurements\cite{dm1}.
 Note also that the upper limit is the only relevant one because if 
$\Omega h^2$ from neutralinos is less than $0.1$ we can always 
imagine some other source of the DM.
\begin{figure}[ht] 
\centerline{ 
\includegraphics[scale=0.45]{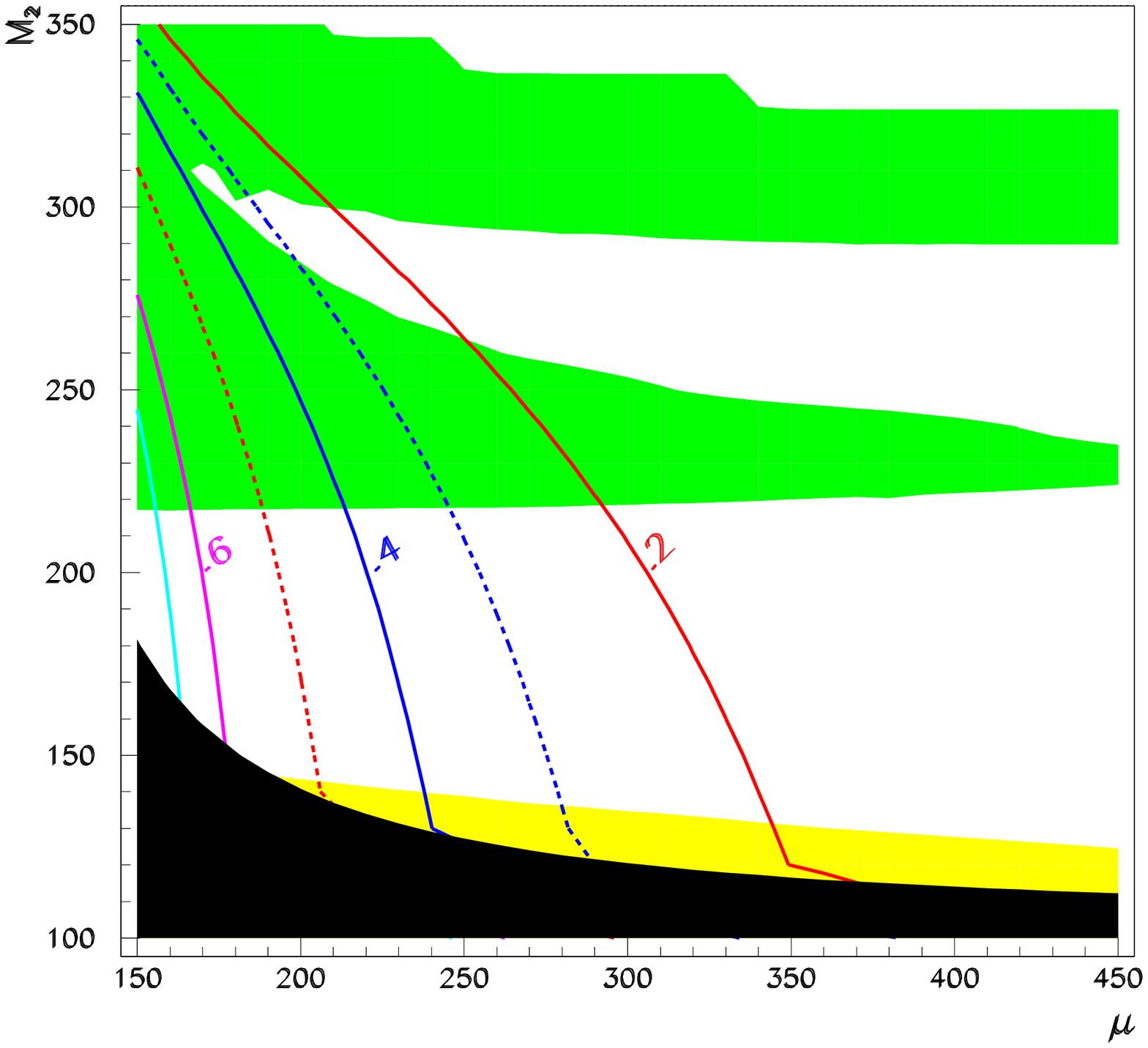} 
\hspace{-1cm}
\includegraphics[scale=0.43]{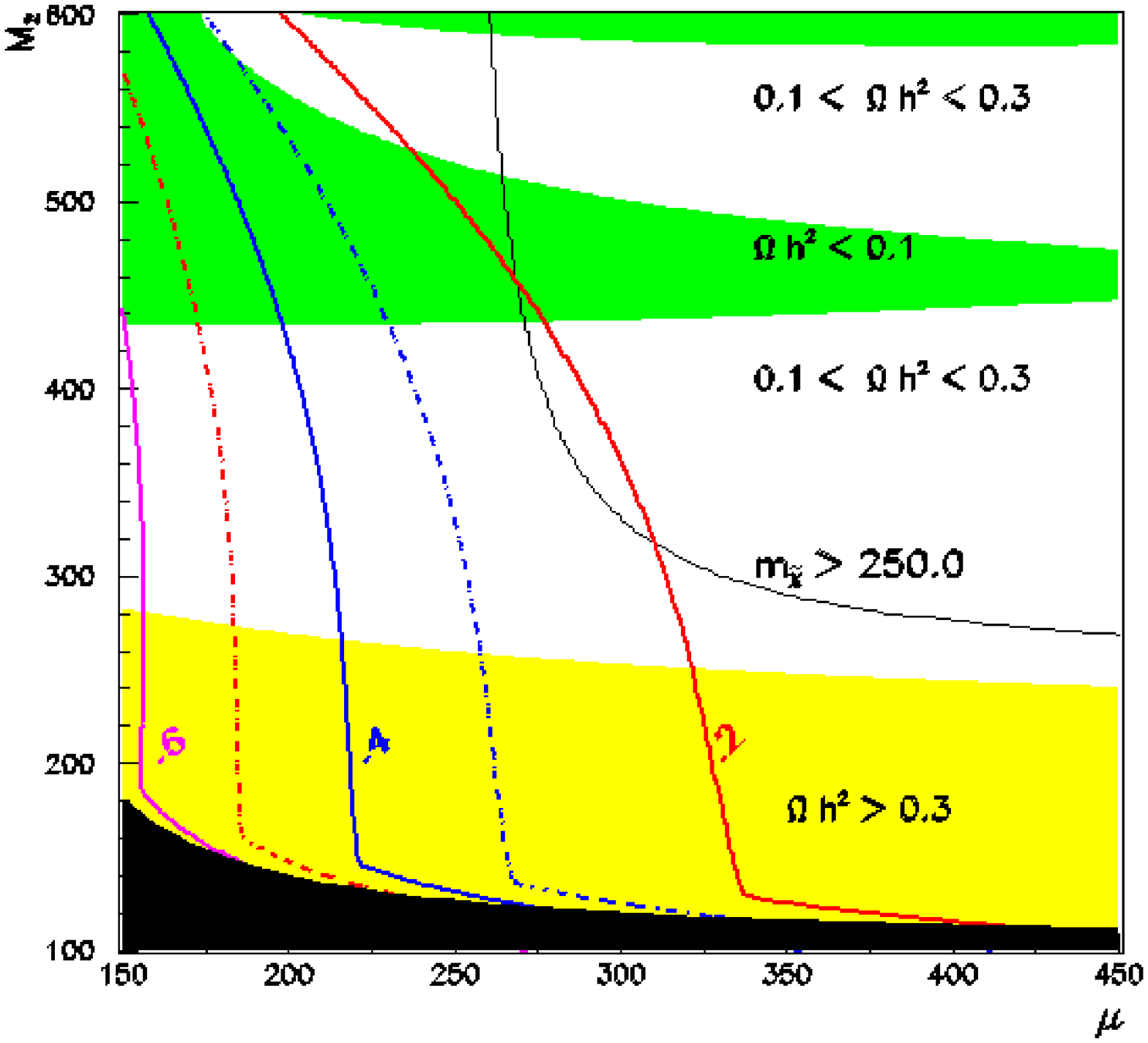}} 
\caption{\label{bmchi510} 
{\em Contours of constant $B.R. (h \ra \neuto \neuto) = .2,.3,.4,.5,.6,(.65)$ 
for $r=.1 (.2)$ in the right (left) panel, along with the
DM  as well as LEP constraints on the $M_2 -\mu$ parameter space. The white 
region is the cosmologically preferred area with $.1 < \Omega h^2 < 0.3$,
for $m_0 = 94 (100) $ GeV. This corresponds to $\tan \beta = 5$ and $m_h = 125$ 
GeV. The black region corresponds to the area excluded by the chargino 
searches at LEP. The lightly (heavily) shaded region corresponds to
$\Omega h^2 > 0.3 (< 0.1)$.}}
\end{figure}
Fig.~\ref{bmchi510} shows in the right (left)  panel contours for B.R. 
$(h \ra \tilde\chi^0_1 \tilde\chi^0_1)$ corresponding to 0.2 to 0.6 (0.65), for
$r = 0.1$ ($r=0.2$), along with regions of different expected values of 
$\Omega h^2$. In this figure we  have used a value of $m_0 = 94(100)$ 
GeV corresponding  to light sleptons and $\tan \beta =5$. 
Due to the efficient annihilation via the $Z/h$ pole one can get regions with 
acceptable relic densities even for heavier slepton masses\cite{belbour2}. 
The panel on the right shows contour for $m_{{\tilde \chi}_1^+} = 250$ 
GeV which indicate the extreme values that could  be probed at the 
Tevatron Run-II. Since in these models the $\tilde \chi_1^0$ 
is lighter than in the ones with universal gaugino masses, the decay products 
of the $\tilde \chi_1^\pm$ should have higher energy. It is obvious 
\begin{enumerate}
\item{There exist regions in the $M_2-\mu$ parameter space where the 
$B.R. (h \ra \neuto \neuto)$ can  be dangerously high to threaten loss of
discovery in both the $\gamma \gamma$ as well as the $b \bar b$ channels 
and which  give rise to acceptable relic density,}
\item{These regions correspond to chargino masses which can be explored 
at the Tevatron.}
\end{enumerate}

We also looked,  by keeping $\tan \beta$ fixed at 5 and scanning over a wide
range of $M_2,M_1$ and $m_0$ values, for the minimum value of $M_1$ 
that one can entertain and  have acceptable relic density.
\begin{figure}[tp]
\begin{center}
\includegraphics[width=16cm,height=10cm]{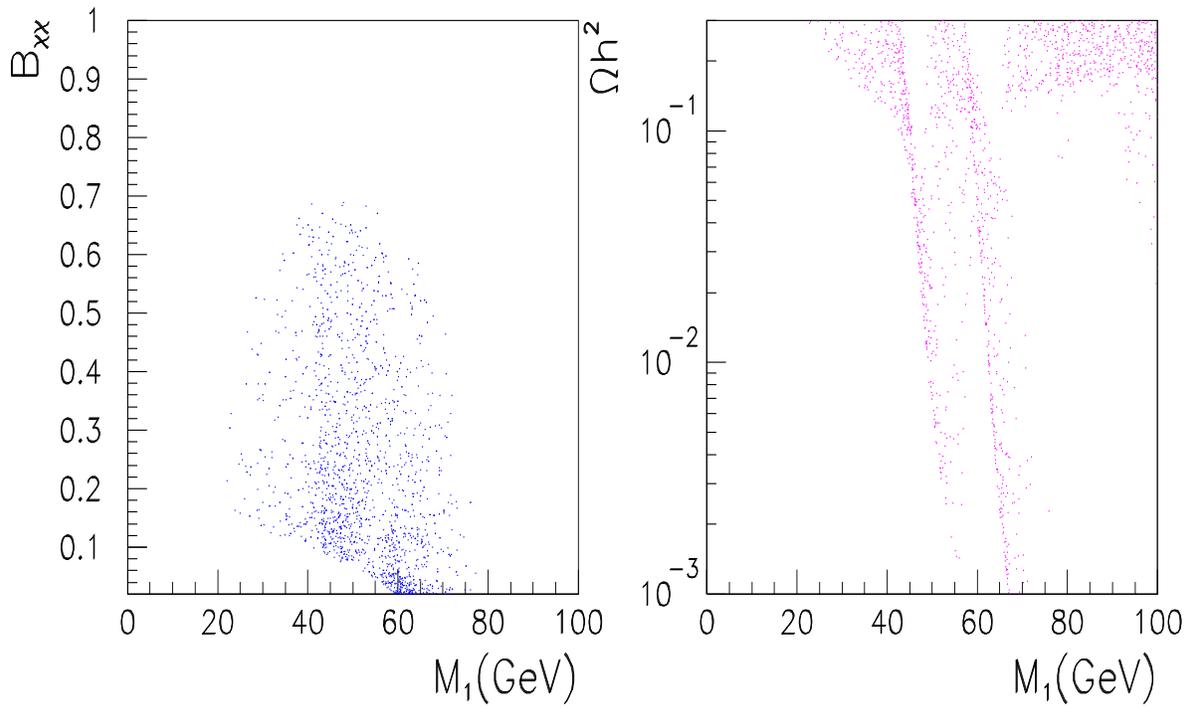}
\caption{\label{tgb5largescan}{\em Large scan over
$M_1,M_2,\mu,m_0$ for $\tgb=5$. The left panel shows the
branching ratio into invisibles {\it vs} $M_1$. The right panel
shows the relic density as a function of $M_1$. Note that one hits
both the $Z$ pole and the Higgs pole. However for the latter
configurations $B_{\chi \chi}$ is negligible.\/}}
\end{center}
\end{figure}
The results are shown in Fig.~\ref{tgb5largescan}.
One sees from this figure that values of $M_1$ smaller than $20$ GeV will lead 
to an unacceptably high relic density, independently of $\mu$, $M_2$ and $m_0$. 
This plot is obtained by a scan over $M_2,\mu$ and $m_0$ values in the range  
$150 < \mu < 500$ GeV, $100 < M_2 < 350$ GeV, $70 < m_0 < 300$ GeV 
and $M_1$ was varied between $10$ and $100$ GeV. The result is also stable with
respect to the variations in $\tan \beta$.

\section{conclusion}
Thus in conclusion we can say the following. It is possible to find substantial
regions in the parameter space  where the `invisible' decay of the lightest
Higgs $h$ into $\neuto$ pairs can dominate in scenarios with nonuniversal
gaugino masses at the high scale. Further we  find that these scenarios
do not  necessarily require a light slepton as they give rise to an acceptable
relic density due to efficient annihilation  at  the $Z$ pole.  The depletion 
into the $\gamma \gamma$ or $b \bar b$ channel can be as low as $0.4$ compared
to the SM.  Such scenarios, do necessarily imply light enough
$\tilde \chi_1^\pm$ and $\tilde \chi_2^0$ which can be
produced at the Tevatron Run-II. However, this also shows the need of
sharpening up the strategies of looking for such an intermediate mass,
`invisibly' decaying Higgs \cite{dp,gunion,wells,zeppen,lesho}.

\noindent
{\bf Acknowledgements} \\
RMG wishes to  thank T. Matsui, Y.Fujii and R.~Yahata for the impeccable 
organisation of the conference in this beautiful place, which provided a
wonderful backdrop for the very nice/useful  discussions that took place. 
She would like to acknowledge financial support of JSPS which made the 
participation possible. Thanks are also due to the LAPTH for their hospitality 
to her for the time when part of this work was carried out.

\end{document}